\documentstyle{aca}

\title{Extinction in the Galaxy from surface brightnesses of ESO-LV galaxies:
determination of $A_R/A_B$ ratio}

\author[J. Cho{\l}oniewski, E. A. Valentijn]
       {Jacek Cho{\l}oniewski,$^1$ Edwin A. Valentijn$^2$\\
       $^1$Astronomical Observatory of Warsaw University,
       Aleje Ujazdowskie 4, PL-00478 Warsaw, Poland,
       e-mail: astro@estymator.com.pl\\
       $^2$Kapteyn Institute, P. O. Box 800, NL-9700 AV Groningen, The Netherlands,
       e-mail: valentyn@astro.rug.nl}

\begin{document}

\maketitle

\begin{abstract}
A new method for the determination of the extinction in our Galaxy is
proposed. The method uses surface brightnesses of external galaxies
in the B and R bands. The observational data have been taken 
from the ESO-LV galaxy catalogue.
As a first application of our model we derive the ratio
of R band to B band extinction. We introduce two methods for computing
the ratio which give: 0.62 $\pm$ 0.05 (the first method) and 0.64 $\pm$
0.06 (the second method) which is in agreement with the recent literature
value of 0.61. 
This agreement confirms the validity and efficiency of our
model and is an independent verification for the standard value 
of the "total to selective extinction".
The method of extinction determination introduced in this paper will be
explored in subsequent publications.

\end{abstract}

\begin{keywords}
dust, extinction - methods: statistical - Galaxy: general - galaxies: fundamental parameters
\end{keywords}

\section{Introduction}
Photometric parameters of external galaxies
have been used several times to derive extinction in our Galaxy. 
Most authors use galaxy colours: 
de Vaucouleurs  \& Buta (1983), Holmberg (1974), Peterson (1970),
Sandage (1973), Teerikorpi (1978). Some of them use galaxy surface
brightnesses: de Vaucouleurs \& Buta (1983), Holmberg (1958)
while the others use galaxy absolute magnitudes: Peterson (1970), Teerikorpi
(1978).

In this paper, we propose a new method of studying extinction
using surface brightnesses of external galaxies in two bands.
We apply this method to surface brightnesses in B and R bands
taken from {\it The Surface Photometry Catalogue of the ESO-Uppsala
Galaxies} (Lauberts \& Valentijn, 1989, hereafter ESO-LV).
The basic idea of the method have been shortly described by Cho{\l}oniewski \&
Valentijn (1991) but the final version of the method described in
this paper is in some important items different.

In comparison with earlier methods which use photometric parameters
of external galaxies (see the beginning of this Section)
our method can be regarded as one which use both surface brightnesses
of galaxies as well as their colours.

The expression for the extinction in the B band as used in
this paper (equation 2) contains a constant, 
$A_R/A_B$, the ratio of the extinction in the R and the B band.
It would be easy to take this ratio from the
literature and use it for extinction determination. But we set out for 
a more ambitious task: we try to extract the $A_R/A_B$ ratio
from ESO-LV surface brightness data alone. An agreement between
our determination and the literature value
would confirm the correctness of our model and
would demonstrate that the surface brightnesses of external galaxies are
a good extinction indicator.

We present here two methods for estimating $A_R/A_B$. Both of
them rely on the simple notion that extinction should not
depend on morphological types of the galaxies, whose surface
brightnesses are used for computing extinction.

\section{The sample}
The ESO-LV galaxy catalogue contains many different surface brightness values for
every galaxy. Three of them are given in two bands (B and R): {\it
surface brightness at half total B light radius} denoted in ESO-LV as 
$\mu_e^B$ and $\mu_e^R$,
{\it average central surface brightness within 10 arcsec diameter circular
aperture} denoted as 
$\mu_0^B$ and $\mu_0^R$ and 
{\it central surface brightness in fit of generalized exponential to 
octants} denoted as $\mu_{oct}^B$ and $\mu_{oct}^R$. 
We use throughout this paper {\it surface brightness
at half total B light} denoted hereafter in this paper as $\mu_B$ and
$\mu_R$. 
We preferred this measure, as the central surface brightness might be
affected by saturation effects and nuclear emission, while the octant
surface brightness represents extrapolated values.
Although the differences of the results obtained using these three
kinds of surface brightness values are small, our analysis
indicates that the surface brightness at half total B light
produce results with the smallest dispersions.

We use in this paper the morphological types of galaxies listed in
the ESO-LV catalogue. However some of them (referred in ESO-LV as
{\it parametrical} ones) are made by means of colour
indicies which, in turn, are extinction dependent. Since it is
vital in this paper to use  morphological types which are
absolutely extinction independent we decided to reject all galaxies with
such morphological types (marked in the ESO-LV catalogue
with $T_{flag}$ equal to 4).

Using any astronomical sample for {\it statistical} 
purposes one has to specify its
definition which clearly describes which objects belong
to the sample. Generally galaxy samples can be apparent diameter-limited
(contain galaxies which have an apparent diameter larger than a certain
diameter limit) or magnitude-limited (contain galaxies which are
brighter than a certain magnitude limit).
The ESO-LV catalogue is diameter-limited: it contains galaxies which have
{\it visual apparent diameter}  in the {\it ESO Quick Blue Survey}
greater or equal to 60 arcsec ($D_{org} \ge 60 \, arcsec$). 

However, this completeness limit depends on morphological type $T$.
But when we take galaxies which have $D_{org} \ge 100 \, arcsec$
the catalogue is complete for all morphological types
(Valentijn 1990, 1994, Huizinga \& van Albada 1992, 
Impey, Bothun \& Malin 1988). 
So we decided to use in this paper galaxies which have
$D_{org} \ge 100 \, arcsec$.

The last (minor) exclusion applied to the original ESO-LV sample was
the rejection of galaxies which have excessive (probably 
wrong) colours: $\mu_B - \mu_R$ smaller than 0 and
greater than 2.1. The exclusion rejects only 18 galaxies.

The final sample, which has been used for the analysis
in this paper contains 2783 galaxies, 19 per cent of the 
full original ESO-LV database (which contains 15467 objects).

\section{The model}

\subsection{The formula for extinction}

As one can see in Fig. 1 the surface brightness in the B band ($\mu_B$) is
strongly linearly dependent on (correlated with) surface brightness in
the R band ($\mu_R$). The amplitude of this linear dependence 
is as high as five magnitudes (similar in both bands) and has
strictly astrophysical origin (it can not be caused by
extinction which is for most galaxies substantially 
less than one magnitude).

The Galactic extinction-free values of $\mu_B$ and $\mu_R$ should, for
a given galaxy, increase 
due to extinction by $A_R$ and $A_B$ respectively (the amount of foreground extinction 
in B and R band respectively).
This means that on the ($\mu_B$,$\mu_R$) plane extinction "moves"
galaxies along the direction with slope $r$:

\begin{equation}
r = \frac{A_R}{A_B}
\end{equation}
which describes the ratio of extinction in R band to the extinction in B
band. 

Taking into account the above facts we may expect that extinction-free
surface brightnesses of galaxies lie on a certain straight line.
The distance to this {\it zero extinction straight line} measured
in the direction with slope $r$ is proportional to the Galactic extinction
(see Fig. 2). 

We test our assumption about the linearity of the {\it zero extinction line}
by fitting straight line to the data on the ($\mu_B$,$\mu_R$) plane
(separately for every morphological type $T$).
Since the data have errors in both coordinates we have used a so called
orthogonal fitting procedure (see Feigelson \& Babu 1992).
We found no systematic residuals from the resulting lines,  confirming
the validity of our linearity assumption.

If we describe the slope of the {\it zero extinction straight
line} as $s^{-1}$ (we use the inverse of $s$ here for symmetry reasons - 
see equations 6, 7, 8, 14 and 15) we can
express extinction in the B band as:

\begin{equation}
A_B = \frac{\mu_B - s \,\, \mu_R}{1 - r \,\, s} \, - \, c
\end{equation}
where $c$ is a certain constant. 

Equation (2) gives a ready-to-use
formula for extinction but in order to use it we have to know
three parameters: $r$, $s$ and $c$. The remaining part of this paper
is mainly devoted just to obtain these.

\subsection{Selection effects}

The sample used in this paper is apparent diameter ($D_{org}$) limited.
We have checked, that the crucial parameters for the present analysis 
: $\mu_B$ and $\mu_R$ are statistically
independent of apparent diameter. 
As we find no dependency
this implies that selection effects do not influence the values of
$\mu_B$ and $\mu_R$, so our formula for $A_B$, which
uses just these two observational values is not
influenced by selection effects (is unbiased).
See Cho{\l}oniewski (1991) for a general discussion
of photometric selection effects.

\subsection{Relation between $r$ and $s$}

Let us define now a new variable $q$ as:

\begin{equation}
q = \mu_R - r \,\, \mu_B .
\end{equation}
This variable (closely related to the Q parameter defined for stars
by Sharpless 1963 in a very similar context) 
is extinction-independent because it is equal to the vertical distance 
on the ($\mu_B$,$\mu_R$) plane between 
a galaxy and {\it extinction direction line}.
As is clearly visible on Fig. 2 the distance to 
{\it the extinction direction line} is insensitive
to extinction because extinction "moves" galaxies on the
($\mu_B$,$\mu_R$) plane in the direction parallel to that line.
So, the variable $q$ should be statistically independent on $A_B$ which
implies that these two variables should be uncorrelated:

\begin{equation}
\varrho ( A_B , q ) = 0
\end{equation}
so their covariance should be equal to zero as well:

\begin{equation}
cov( A_B , q ) = 0.
\end{equation}
When we replace in equation (5) $A_B$ and $q$ by expressions
given in equations (2) and (3) respectively and apply the
law of error propagation (Brandt 1970) we obtain a relation
which binds the variables $r$ and $s$:

\begin{equation}
r \, \sigma^2(\mu_B) \, + \, s \, \sigma^2(\mu_R) = (1 \, + r \, s)  \, 
cov(\mu_B,\mu_R)
\end{equation}
where $\sigma^2$ denotes the square of the standard deviation (variance).
Equation (6) defines on the ($\mu_B$,$\mu_R$) plane a hyperbola 
(see Fig. 3) and may be solved with respect to $s$:

\begin{equation}
s\,=\,\frac{r\,\sigma^2(\mu_B) - cov(\mu_B,\mu_R)}
           {r\,cov(\mu_B,\mu_R) - \sigma^2(\mu_R)}
\end{equation}
and $r$:
\begin{equation}
r\,=\,\frac{s\,\sigma^2(\mu_R) - cov(\mu_B,\mu_R)}
           {s\,cov(\mu_B,\mu_R) - \sigma^2(\mu_B)} \, .
\end{equation}

\subsection{Determination of $c$}

Our method can provide only relative extinction, namely: extinction
plus an unknown constant. So, we are free to normalize 
our extinction estimate to have:  

\begin{equation}
< A_B > \, = \, 0 \, ,
\end{equation}
what means that our extinction is the relative extinction compared to
an overall mean. Equation (9) applied to equation (2) gives an
expression for the constant $c$:

\begin{equation}
c \, = \, \frac{< \mu_B > \, - \, s \, < \mu_R >}
               {1 \, - r \,\, s}.
\end{equation}

\subsection{How to compute?}

The average values (more precisely: the expectation values) 
present in equations (9) and (10) can be easily
computed using the well known formula:

\begin{equation}
< x > \, = \, \frac{\sum_{i=1}^{n} x_i}{n}
\end{equation}
where summation is done over the data in the galaxy sample.
The standard deviations and covariances present in most equations in this paper ( can be computed from the 
following formulae:

\begin{equation}
\sigma^2(x) \, = \, <x^2> \, - \, <x>^2
\end{equation}

\begin{equation}
cov(x,y) \, = \, <x\,y> \, - \, <x>\,<y>
\end{equation}
where $x$ and $y$ denotes any variable used in this paper.

\subsection{Statistical and geometrical context}

Equations (2) and (3) define in fact a linear
transformation of the "old" variables $\mu_B$ and $\mu_R$
into the "new" variables $A_B$ and $q$. The transformation
is defined in such a way that the "new" variables have
to be uncorrelated. The coefficients of such a linear
transformation are bound by equation (6). 

We solve here a general problem: how to transform linear "old"
variables $x$ and $y$ into the "new" ones $X$ and
$Y$:

\begin{equation}
X \, = x \, - \, s \, y
\end{equation}
\begin{equation}
Y \, = y \, - \, r \, x
\end{equation}
while $X$ and $Y$ are uncorrelated:

\begin{equation}
\varrho(X,Y) \, = \, 0
\end{equation}
The solution of this problem, lies in the condition
for $s$ and $r$:
\begin{equation}
r \, \sigma^2(X) \, + \, s \, \sigma^2(Y) = (1 \, + r \, s)  \, 
cov(X,Y)
\end{equation}
Equation (6) is of course the special case of this equation.

Geometrically, the linear transformation of two "old"
variables into the "new" ones is equivalent to applying
a certain new coordinate system, which uses two "new" axes:
the $X$ "new" axis is tilted to the "old" $x$ axis
by a slope $r$, while the "new" $Y$ axis is tilted to the
"old" $y$ axis by a slope $s$.

Our solution is closely related to the procedure which 
leads to the so called {\it orthogonal regression line}
(see Feigelson \& Babu 1992). Such a line is constructed
by rotating coordinates in a way to obtain
uncorrelated data in "new" coordinates. 
Our solution can be regarded as a generalization of 
this procedure: we can easily reconstruct an
{\it orthogonal regression line} by substituting
in equation (17) $s = -r$. 

\section{Determination of $A_R/A_B$: first method}

We compute in this Section the ratio of extinction
in R band to the extinction in B band denoted
as $r$ (see equation 1) using the simple fact that
extinction $A_B$ must be statistically independent on galaxy morphological
type $T$.

\subsection{Mathematical approach}

The statistical independence of $A_B$ and $T$ imply that:

\begin{equation}
\varrho(A_B,T) \, = \, 0
\end{equation}
which gives:

\begin{equation}
cov(A_B,T) \, = \, 0 .
\end{equation}
Taking into account equation (2), using the law of error propagation
(Brandt 1970) and assuming that the parameters
$s$ and $c$ do not depend on $T$ we have:

\begin{equation}
s \, = \, \frac{cov(\mu_B,T)}{cov(\mu_R,T)} .
\end{equation}
This value for $s$ substituted into equation (8) gives the formula for $r$:

\begin{equation}
r\,=\,\frac{\frac{cov(\mu_B,T)}{cov(\mu_R,T)}\,\sigma^2(\mu_R) - cov(\mu_B,\mu_R)}
           {\frac{cov(\mu_B,T)}{cov(\mu_R,T)}\,cov(\mu_B,\mu_R) -
\sigma^2(\mu_B)} \, .
\end{equation}

Equations (20) and (21) hold under the assumption that $s$ and $c$ do not
depend on $T$. 
Let us check now whether this is true by
computing $s$ and $c$, for different
values of $r$ using equations (7) and (10).
The uncertainties for $s$ and $c$ have been computed using a so
called jacknife method (see Quenouille 1956, Tukey 1958 and
Efron \& Tibshirani 1986).
The results are given on Fig. 4 and 5.  It is evident that
$s$ and $c$ are $T$ dependent, but for morphological type $T$
between 2.5 and 6.5, both parameters are constant for any value
of $r$ in the limits of accuracy achieved here. 

For this reason we computed $r$ using equation (21) 
taking only those galaxies for with $T$ is greater than
2.5 and less than 6.5. 
The resultant subsample contains 1290 objects
(approximately half of the whole sample). 
We computed for these galaxies all the components of the equation (21):

\begin{equation} cov(\mu_B,T) = 279.032         \end{equation}
\begin{equation} cov(\mu_R,T) = 416.098         \end{equation}
\begin{equation} \sigma^2(\mu_B) = 0.432634     \end{equation}
\begin{equation} \sigma^2(\mu_R) = 0.530865     \end{equation}
\begin{equation} cov(\mu_B,\mu_R = 0.440979     \end{equation}
What finally gives:

\begin{equation}
r = 0.62 \pm 0.05
\end{equation}
where the uncertainty of the result was again computed using the
jacknife method.

\subsection{Graphical approach}

The idea which has been used here for computing $r$ ($A_B$ does
not depend on $T$) can be applied in another way. Let us compute
the average extinction (according to equations 2 and 7) 
and plot it as a function of morphological type
(denoted as $<A_B>_T$) for different values of $r$ - see Fig. 6.
It is clearly visible that only for $r=0.6$ the extinction 
$A_B$ is not 
a function of morphological type $T$. So we arrive again at a
similar result as expressed in equation (21).

\section{Determination of $A_R/A_B$: second method}

We compute in this Section the ratio of extinction in R band to the
extinction in B band denoted as $r$ (see equation 1) using the
fact that the standard deviation of extinction $A_B$ must
be the same for every galaxy morphological type $T$.

According to equation (2) and applying the law of error propagation
(Brandt 1970) the value of $\sigma^2(A_B)$ can be expressed
as:

\begin{equation}
\sigma^2(A_B)\,=\,\frac{\sigma^2(\mu_B)+s^2\sigma(\mu_R)-2\,s\,cov(\mu_B,\mu_R)}
                       {(1\,-\,r\,s)^2}
\end{equation}
When we substitute $s$ present in equation (28) by expression 
given in equation (7) the resultant
equation for $\sigma^2(A_B)$ depends only on $r$. 
When we use it for
two different morphological types (say $T_1$ and $T_2$) we have:

\begin{equation}
\sigma_{T_1}(A_B)_r \, = \, \sigma_{T_2}(A_B)_r
\end{equation}
which can be solved with respect to $r$. Such solution is presented
graphically on Fig. 7 for $T_1=-3$ and $T_2=5$. The appropriate
two curves intersect at $r=0.66$ which is the solution for $r$ for 
this particular pair of $T$. 

The values of $\sigma_T(A_B)_r$ as a function of $r$ for all morphological types 
($T$ = -5, ... , 10) are presented in Fig. 8. The curves intersect in
many different points so we have many different "solutions" for $r$. 
This motivated us to compute the relative standard deviation of
$\sigma_T(A_B)_r$, namely:

\begin{equation}
D(r)\,=\,\frac{\sigma(\sigma_T(A_B)_r)}{<\sigma_T(A_B)_r>} .
\end{equation}
The minimum of the function $D(r)$ will be the solution
for $r$.
For two morphological types: $T$=-5 and $T$=9
$\sigma_T(A_B)_r$ strongly differs from the mean so 
we decided to omit these two types in our final computations in
this Section (the rejected curves are marked on Fig. 8 by a dotted
lines).

The reason for existence of such two "outliers" is probably that
thay have different intrinsic extinction-free scatter with
respect to the "zero extinction" straight line on the 
($\mu_B$,$\mu_R$) plane than for other galaxies: 
this scatter is smaller (in comparison with
most galaxies) for $T$=-5 and greater for $T$=9. 

The function $D(r)$ has a minimum for:

\begin{equation}
r\,=\,0.64 \pm 0.06
\end{equation}
where the uncertainty has been computed by the jacknife method
(with respect to $T$).

\section{Discussion}

The second method for determination of the $A_R/A_B$ ratio
works only due to the fact
that the parameter $s$ is $T$ dependent. Moreover, 
the stronger this dependence is, the more accurate are the results given
by the second method. The opposite situation is for the first method:
it works only for such galaxies for which the $s$ parameter is {\it not}
$T$ dependent. 
So the first method can survive only due to galaxies 
with the same $s$ parameter
which is in contradiction to the second method which 
feeds itself by the differences in $s$.
While complementary in nature, both tests gives similar results.

Both methods have their specific disadvantages. The first one
does not use the whole sample. The second one relies on the 
assumption that the extinction-free scatter of galaxy 
surface brightnesses is the same for every morphological 
type (what is not always true).

The value for $A_R/A_B$ which was recently published
is 0.61 (Schlegel, Finkbeiner \& Davis 1998). It takes
into account several observational and instrumental factors
and is based on formulae published by Cardelli, Clayton \& Mathis 1989 and
O'Donnell 1994. Our results: 0.62 $\pm$ 0.05 (the first method)
and 0.64 $\pm$ 0.06 (the second method) are in very good agreement
with this value. It confirms the correctness
of the model introduced in this paper and practically
demonstrates that two band surface brightness data of external
galaxies are a good extinction indicator.

Since $A_R/A_B$ ratio is closely
connected to "total to selective extinction ratio" $A_V/E(B-V)$ our
result confirms its standard value which is equal to 3.1.

The method applied in this paper to obtain $A_R/A_B$ ratio 
relies on the assumption that extinction does not depend on
morphological type. In future applications to other data
(e.g. SDSS) one can use, instead of morphological type, other extinction
independent quantities like: axial ratio, redshift or effective
diameter.

\section*{Ackonwledgments}

The part of this work was done during the stay of the authors
at European Southern Observatory (Garching, Germany).
The authors thanks Andris Lauberts for
collaboration about the project.

\clearpage
\begin{figure*}
\vspace{220mm}
 \includegraphics{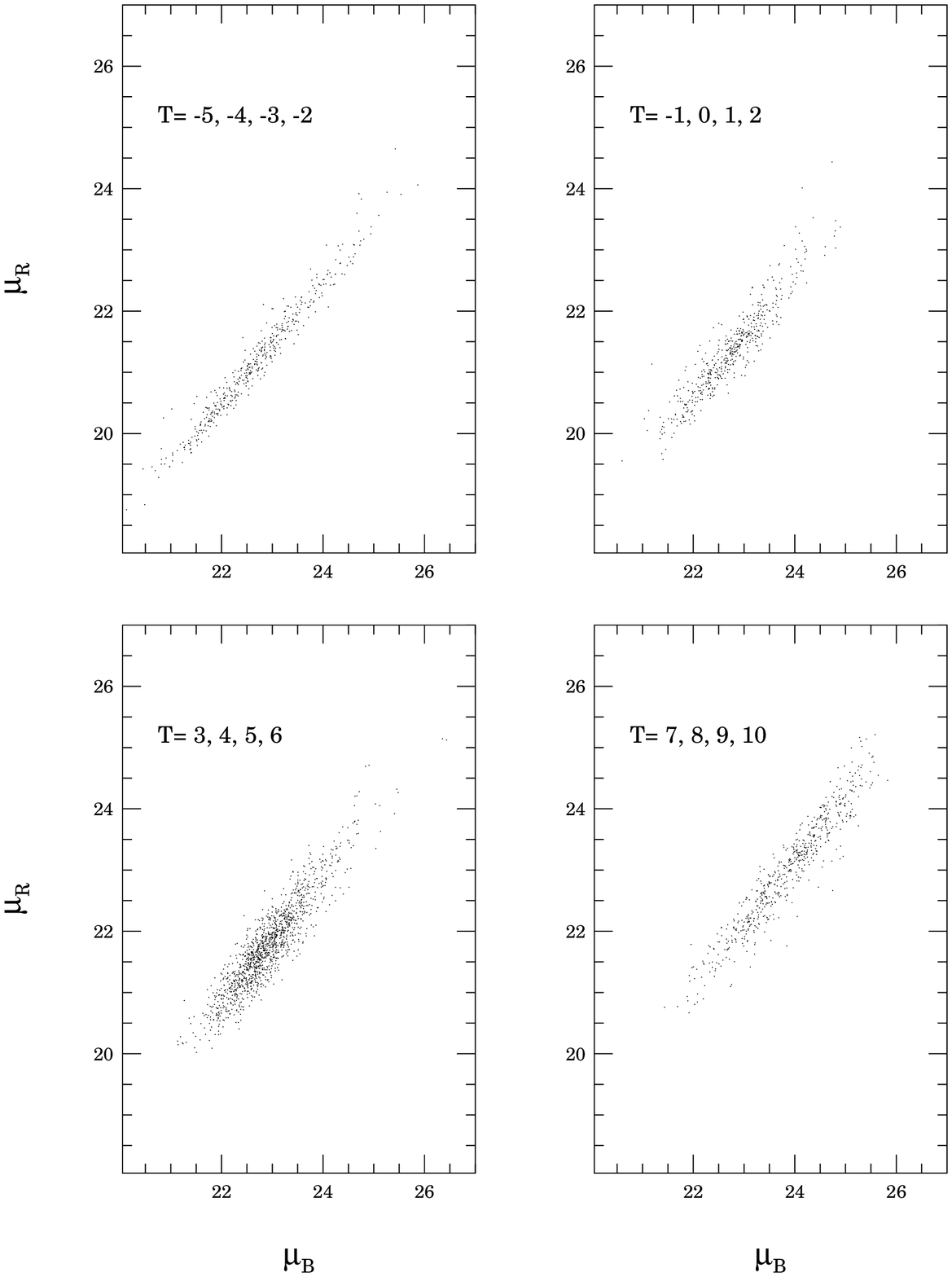}
 \caption{
Surface brightnesses at half total B light radius 
of galaxies in the B and R bands for four
different morphological type $T$ intervals. The data are from
a subsample of the ESO/LV galaxy catalogue which has been used for
all the computations presented in this paper.
}
\end{figure*}

\clearpage
\begin{figure*}
\vspace{220mm}
 \includegraphics{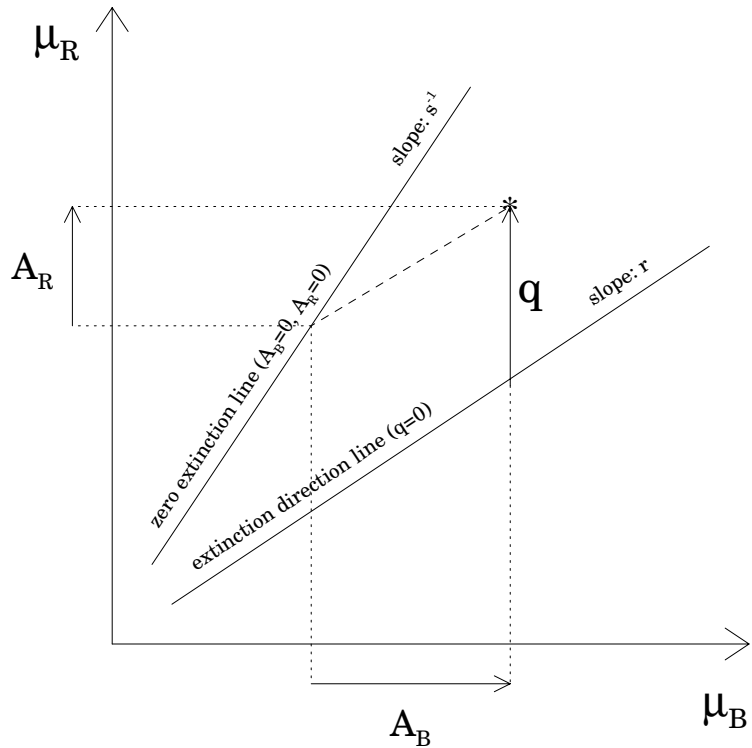}
 \caption{
The idea of extinction measurements using surface brightness of galaxies
in two bands: $\mu_B$ and $\mu_R$. The distance of a galaxy 
(marked as an asterix) to the {\it zero extinction line}
measured parallel to the {\it extinction direction line} is
proportional to the Galactic extinction. The $q$ parameter is extinction
independent.
The slope of the {\it zero extinction line} is $s^{-1}$ while the slope
of the {\it extinction direction line} is $r=A_R/A_B$.
}
\end{figure*}

\clearpage
\begin{figure*}
\vspace{220mm}
 \includegraphics{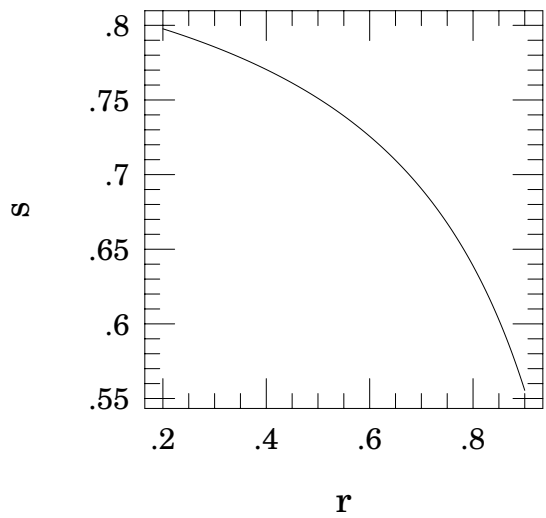}
 \caption{
The relation between the parameters $s$ and $r=A_R/A_B$ 
derived from the requirement
that extinction $A_B$ and the parameter $q$ are statistically
independent (as described in equation 6).
}
\end{figure*}

\clearpage
\begin{figure*}
\vspace{220mm}
 \includegraphics{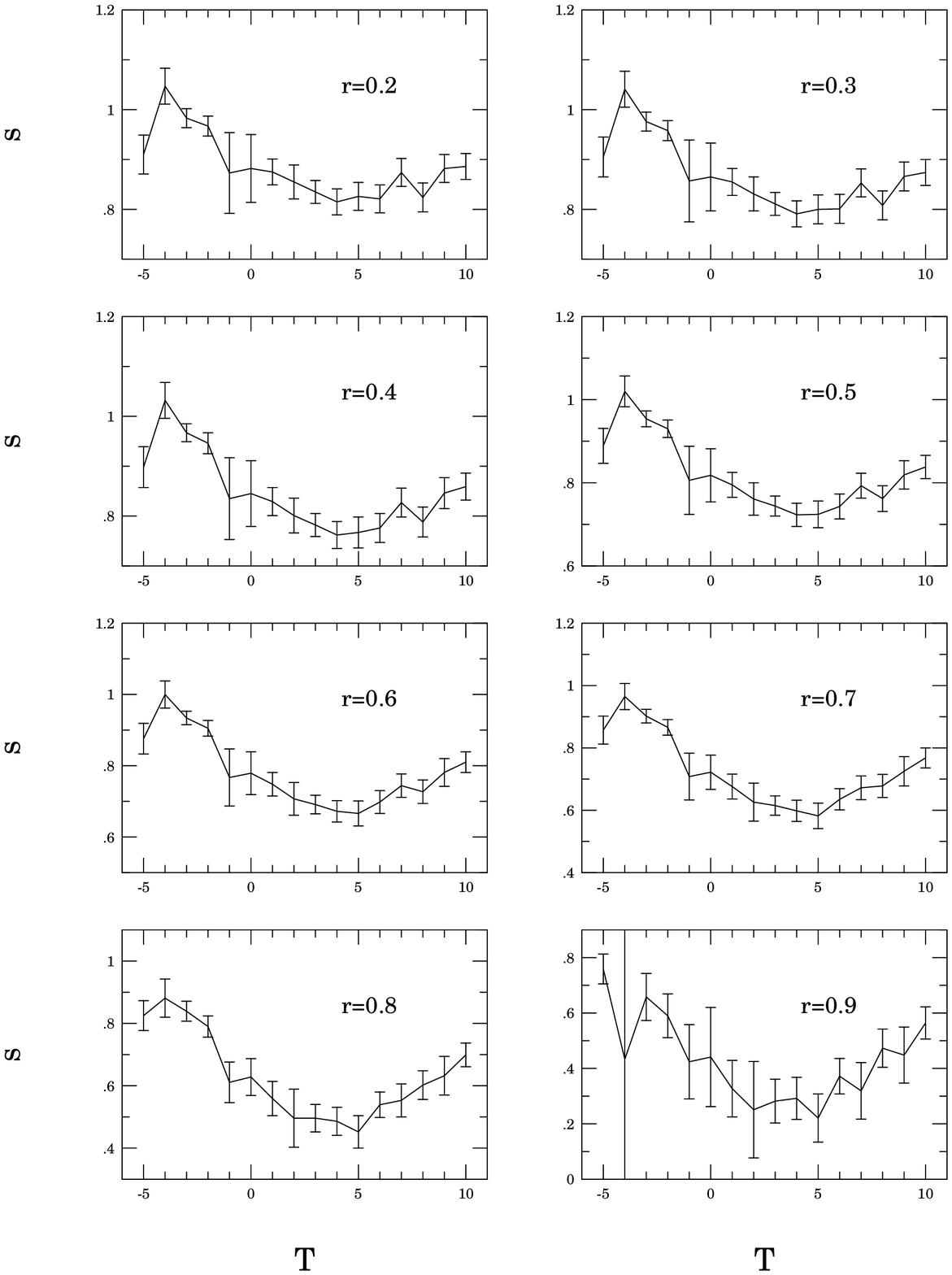}
 \caption{
The parameter $s$ as a function of morphological type $T$ for various
values of $r=A_R/A_B$. 
Error bars represent standard deviation ($1\sigma$).
}
\end{figure*}

\clearpage
\begin{figure*}
\vspace{220mm}
 \includegraphics{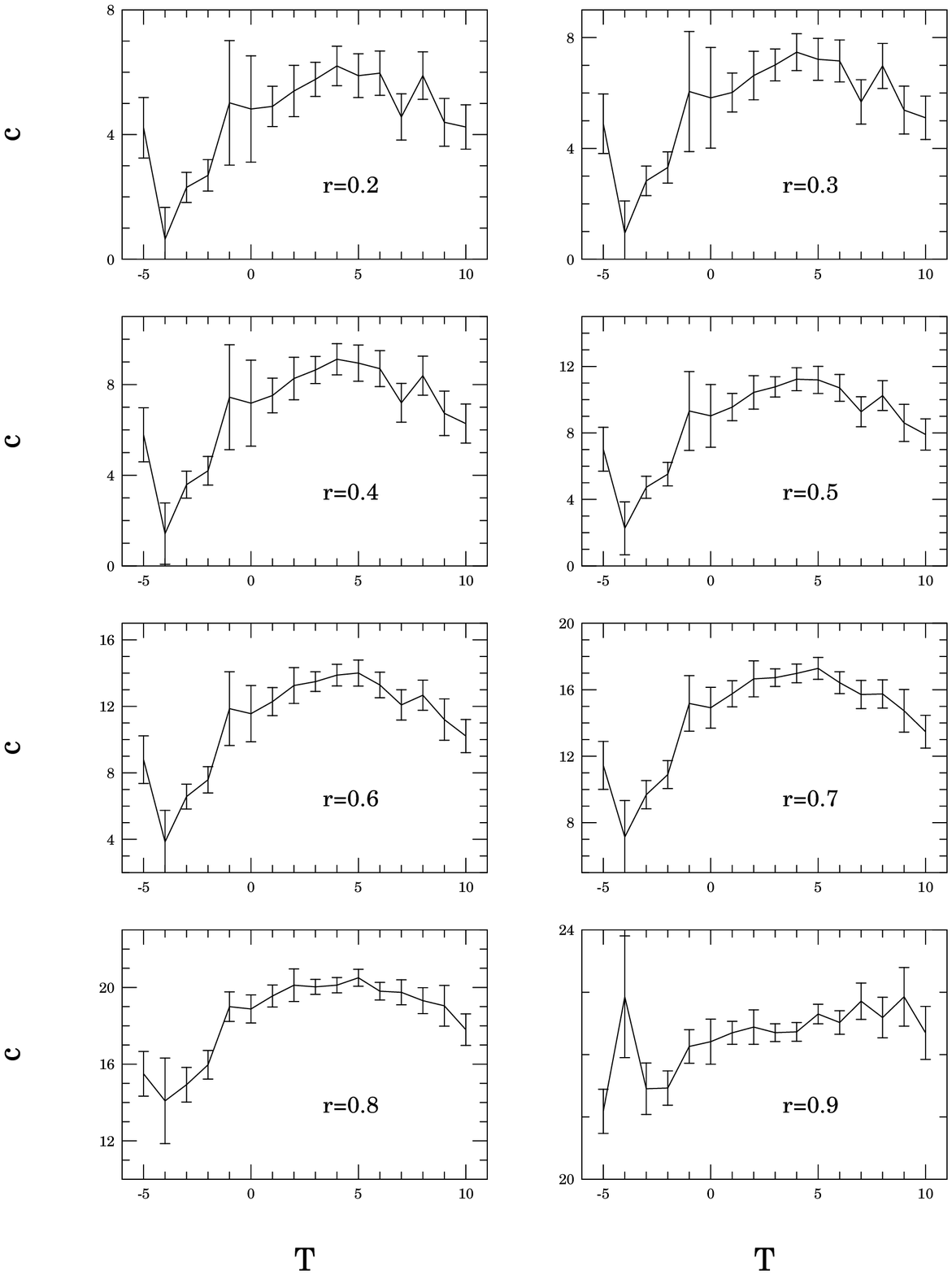}
 \caption{
The parameter $c$ as a function of morphological type $T$ for various
values of $r=A_R/A_B$.
Error bars represent standard deviation ($1\sigma$).
}
\end{figure*}

\clearpage
\begin{figure*}
\vspace{220mm}
 \includegraphics{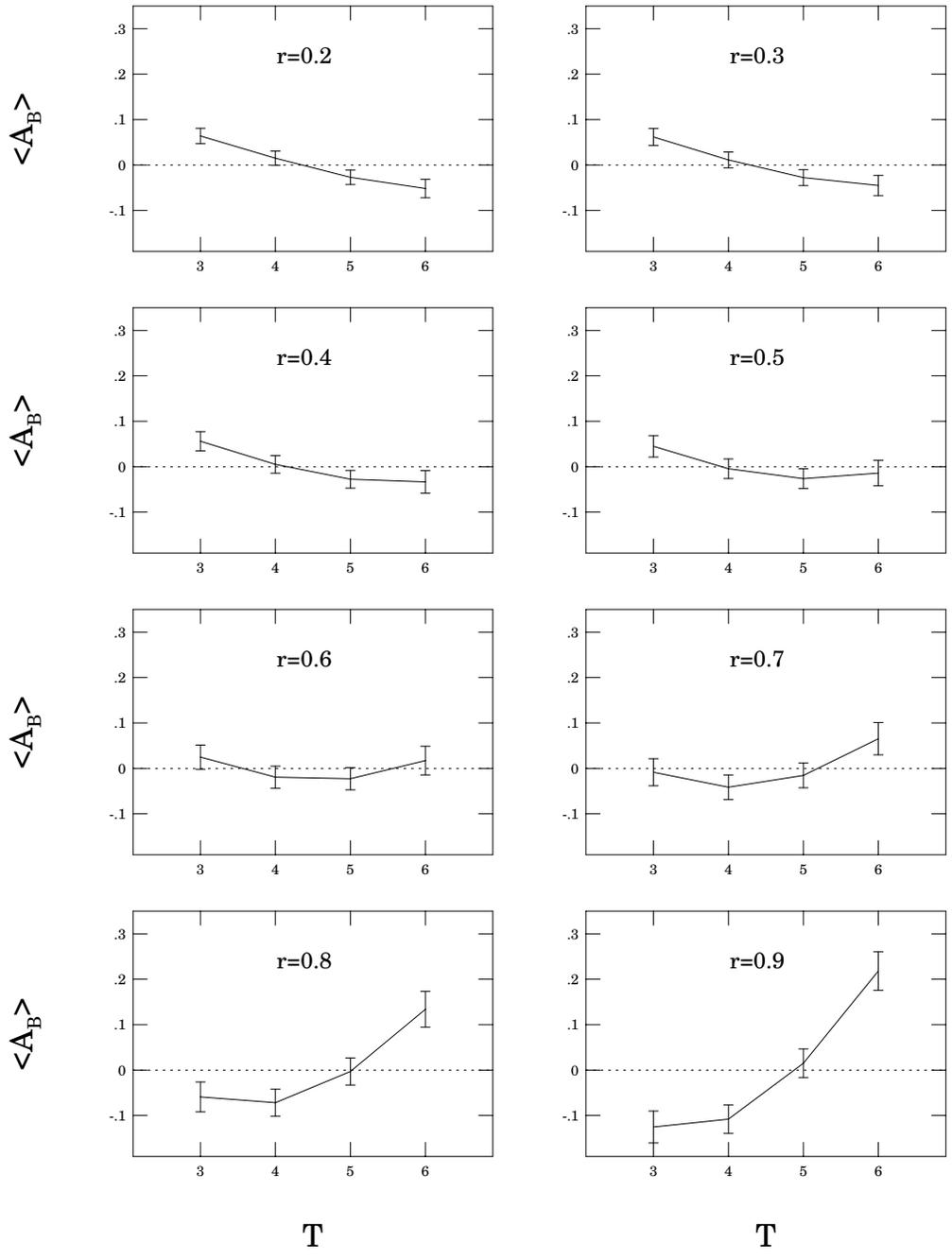}
 \caption{
The average extinction in B band $<A_B>$ as a function of morphological type $T$
for various values of $r=A_R/A_B$. The extinction should not
depend on morphological type which is true only for $r$=0.6.
Error bars represent standard deviation ($1\sigma$).
}
\end{figure*}

\clearpage
\begin{figure*}
\vspace{220mm}
 \includegraphics{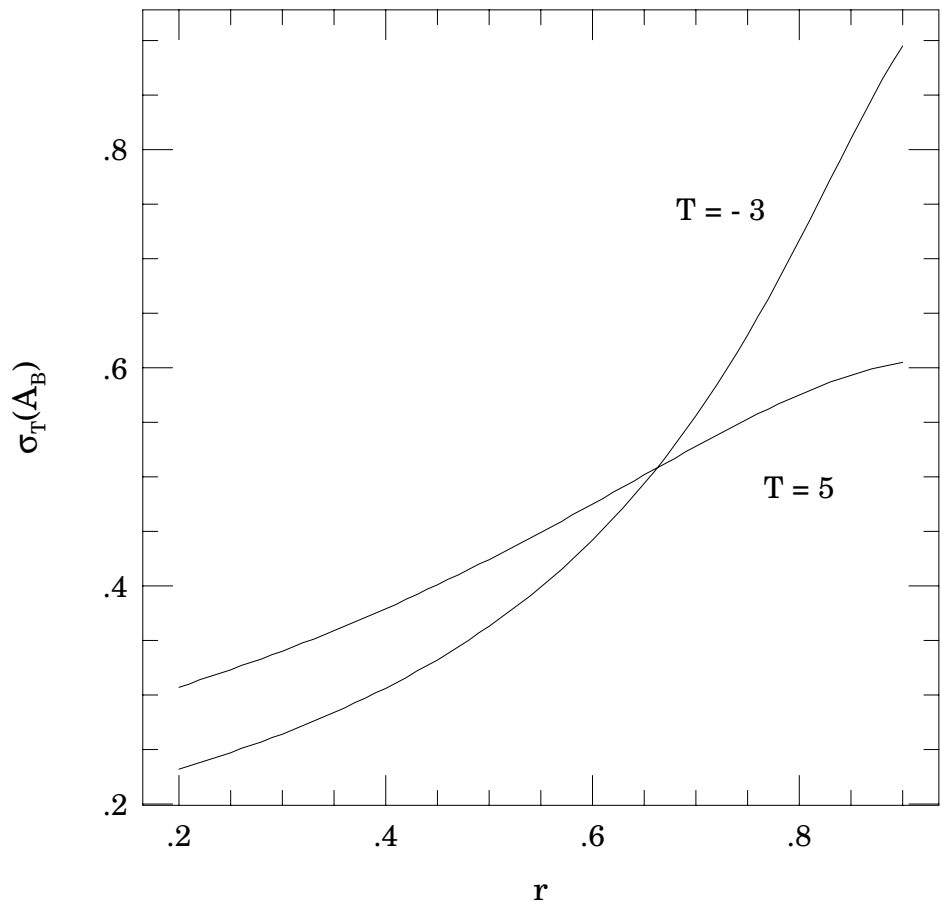}
 \caption{
The standard deviation of extinction in B band $\sigma(A_B)$ as
a function of $r=A_R/A_B$ for galaxies with $T=-3$ and $T=5$. 
The point of intersection of these two curves at
$r$=0.66 gives the solution for $A_R/A_B$ since $\sigma(A_B)$
should be the same for any $T$.
}
\end{figure*}

\clearpage
\begin{figure*}
\vspace{220mm}
 \includegraphics{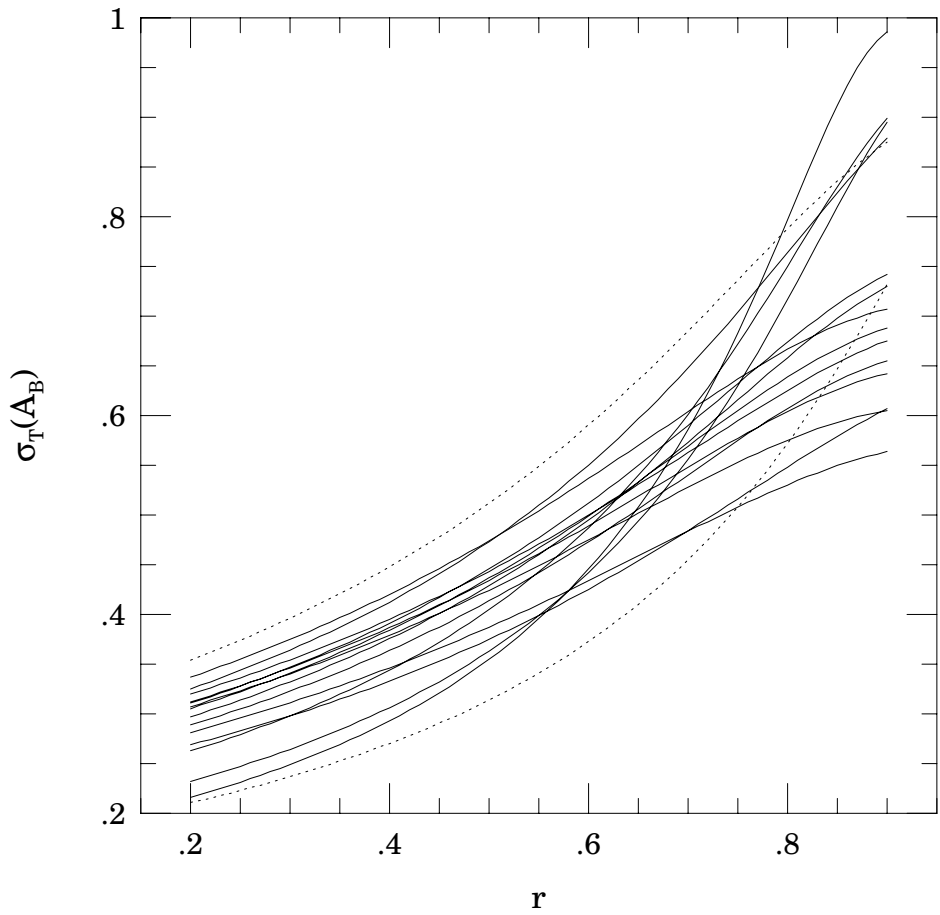}
 \caption{
The same as Fig. 7 but for all morphological types $T$= -5, ... , 10).
The dotted line refers to galaxies with $T$=-5 (at the bottom) and $T$=9
(at the top). These galaxies strongly differ from the mean so they
have been rejected in final computations.
}
\end{figure*}

\clearpage
\begin{figure*}
\vspace{220mm}
 \includegraphics{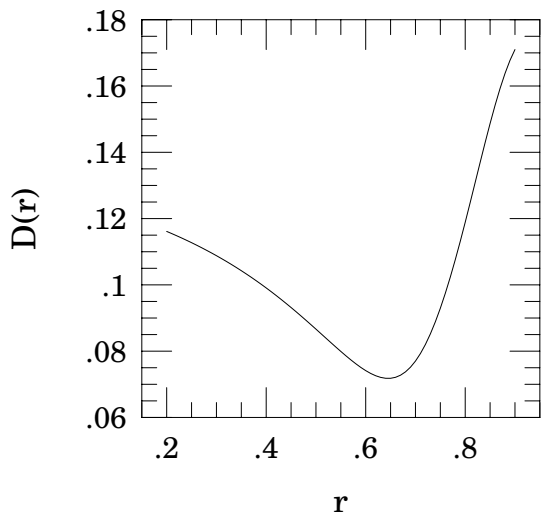}
 \caption{
The function $D(r)$ which describes the relative standard deviation
of the $\sigma(A_B)$. 
The minimum of this function for $r$=0.64 $\pm$ 0.06 represents the
solution for $r=A_R/A_B$ since $\sigma(A_B)$ should be
as close as possible to each other (in ideal case the same)
for all morphological types.
}
\end{figure*}

\end{document}